# Enhanced Magnetic Anisotropy of $Mn_{12}$-acetate

D.M. Seo[a], V. Meenakshi[a], W. Teizer[a] [*], H. Zhao[b], and K. R. Dunbar[b]

[a]Department of Physics, Texas A&M University, College Station, TX 77843- 4242, USA

[b]Department of Chemistry, Texas A&M University, College Station, TX 77842-3012, USA

*Corresponding author: (Tel) 1-979-845-7730, (Fax) 1-979-845-2590, (e-mail) teizer@tamu.edu

## Abstract

Thin films of the Single Molecule Magnet $[Mn_{12}O_{12}(CH_3COO)_{16}(H_2O)_4]\cdot 2CH_3COOH\cdot 4H_2O$ ($Mn_{12}$-acetate) have been fabricated on a Si-substrate by the Dip-and-Dry method, a simple and robust technique. Atomic force microscopy and X-ray photoelectron spectroscopy characterizations reveal that homogeneous, thin films of a few molecular layers with smoothness at the molecular level are deposited. Significant changes in magnetic properties of $Mn_{12}$-acetate exposed to the same solvent were observed in zero-field-cooled and field-cooled magnetization, as well as ac-susceptibility measurements. The blocking temperature was found to increase to $T_B > 10$ K at low magnetic fields, indicating an enhanced magnetic anisotropy.

**Keywords:** $Mn_{12}$-acetate; Thin film; Blocking Temperature; Magnetic Anisotropy

**PACS:** 75.70.Ak

# 1. Introduction

Single molecule magnets (SMMs), the most studied of which is $[Mn_{12}O_{12}(CH_3COO)_{16}(H_2O)_4]\cdot 2CH_3COOH\cdot 4H_2O$ ($Mn_{12}$-acetate) [1-4], provide a model system for the study of quantum tunneling of the magnetization [3]. Stepwise magnetization hysteresis loops and out-of-phase ac susceptibility signals are due to a high spin ground state and a strong uniaxial magnetic anisotropy of $Mn_{12}$-acetate [2], in which eight $Mn^{3+}$ (S=2) ions and four $Mn^{4+}$ (S=3/2) ions are magnetically coupled by oxygen bridges to form the S=10 ground state [5]. Moreover, these compounds have also been considered for future applications such as quantum computing and information storage devices [6,7]. The interesting magnetic properties in these materials arise from individual molecules rather than intermolecular interactions. For this reason, single molecules can, in principle, be used to store magnetic information. In order to use these molecules in devices, however, an increase of the blocking temperature ($T_B$) of these materials is essential.

There are few reports on the film production of $Mn_{12}$ derivatives to date [8-14], and even less reports on the significant enhancement of magnetic properties from that of the parent compound [11,12]. Here we report the production of thin, homogeneous $Mn_{12}$-acetate films of continuous coverage by the dip-and-dry (DAD) method, and the enhancement of the magnetic anisotropy of $Mn_{12}$-acetate, which has undergone similar solvent exposure. Two studies regarding the magnetic properties of $Mn_{12}$ derivatives formed by thermal transformation and gas inclusion as well as incorporation into mesoporous silica have been reported [15,16]. In the present study, an increase of $T_B$ to > 10 K in dc-susceptibility measurements was observed, a significant change from the parent compound behavior that shows a zero-field blocking temperature of ~ 3.5 K, as observed in reference [3] and in our as-produced $Mn_{12}$-acetate powder. These observations are of fundamental importance and a promising first step for potential applications of these materials. Atomic force microscopy (AFM) was used to investigate the surface morphology and the thickness of the films, which revealed roughness on the molecular scale and a thickness of ~1 molecular layer per dip. X-ray photoelectron spectroscopy (XPS) measurements were also carried out to analyze the electronic structure of the thin films.

## 2. Experimental

A fresh sample of $Mn_{12}$-acetate was prepared following the customary procedure [1]. For a typical preparation of films described below by the DAD technique, 2.2 mg ± 0.1 mg of $Mn_{12}$-acetate was dissolved in 10 ml of acetonitrile ($CH_3CN$) to produce a $1.1 \times 10^{-4}$ mol·L$^{-1}$ solution. Prior to the DAD step, the $Si/SiO_2$ substrate was rinsed with acetone and isopropanol. The clean wafer was dipped in the prepared $Mn_{12}$-acetate solution and immediately removed. A thin film of the solution was subsequently observed on the substrate, which dried within several seconds to produce a thin film of $Mn_{12}$-acetate. All procedures were carried out inside a fume hood under ambient conditions.

After preparation of thin films by the DAD technique, the surface morphology of the films was studied by AFM, with a Digital Instruments Nanoscope IIIa. The AFM images were acquired in the tapping mode with a silicon cantilever and tip under ambient conditions. Room temperature core level XPS measurements were performed using a Kratos AXIS ULTRA spectrometer equipped with a concentric hemispherical analyzer using the Al Kα radiation (hv=1486.6 eV) and a base pressure of ~ $2 \times 10^{-8}$ Torr. The binding energies were calibrated with respect to the C 1s peak (284.8 eV).

## 3. Results and discussion

### 3.1. AFM and XPS characterizations

Fig. 1 shows AFM images of the $Mn_{12}$-acetate thin film. A topographical top-view, the corresponding 3D image, and height profile are shown for $1 \times 1$ μm$^2$ scan size, respectively from (a) to (c). This figure shows a $Mn_{12}$-acetate thin film, which, considering the simplicity of the DAD method, is surprisingly homogeneous and smooth. In a control experiment, the DAD method was used with pure acetonitrile instead of the $Mn_{12}$-acetate solution. As expected, the resulting images (not shown) did not show any substantial surface corrugations. Detailed examination of the AFM images (Fig. 1) reveal that the typical horizontal size of pictured $Mn_{12}$-acetate clusters is about 25 nm and the average vertical height is about 2 nm, close to a molecular diameter. As a result of the radius of curvature of an AFM tip (~ 10 nm) [17], the typical horizontal size of a cluster appears larger than a single molecule. The height information, however, indicates that nearly all of the particles on the film form a monolayer instead of clusters. The root mean square (RMS)-roughness of the surface is 0.73 nm, considerably smaller than the size of a single molecule (1.7 nm) [1].

The thickness of the films after a single DAD step was measured and found to be ~ 2 nm in AFM measurements of artificial $Mn_{12}$-acetate patterns (not shown). The $Mn_{12}$-acetate films were surprisingly homogeneous, both microscopically and macroscopically, as indicated by the fact that all the AFM images we have acquired in different regions of the film show similar images [18].

In order to analyze the electronic structure of the thin films after the DAD process, X-ray photoelectron spectroscopy (XPS) measurements have been carried out. Multiple peaks of C 1s, and O 1s, which show the existence of different C and O sites in $Mn_{12}$-acetate [19], were observed in narrow scan spectra with pass energy of 40 eV (not shown). Core-level spectra of Mn 2p for the crystalline $Mn_{12}$-acetate (pellet) and the $Mn_{12}$-acetate thin film, which was used for AFM, are shown in Fig. 2. The two peaks for the Mn 2p core-level of $Mn_{12}$-acetate thin films, at 642 eV and 653.6 eV, correspond to $2p_{3/2}$ and $2p_{1/2}$, respectively. A small shift, observed in the binding energy for different samples, was within the XPS resolution (0.5 eV). The films are thin enough to observe the Si 2p peak from the substrate, at 99.5 eV and 103.1 eV, that correspond to a pure silicon phase and a silicon oxide phase, respectively (not shown) [20].

## 3.2. Magnetic characterizations

We were not successful in acquiring magnetization data using these DAD samples due to the small amount of material and the background signal from the substrate. Therefore, a sample for the magnetization measurements was made by a similar method to the DAD samples but on a larger scale to accumulate sufficient sample material. A solution of ~ $1 \times 10^{-3}$ mol·L$^{-1}$ (21 mg of $Mn_{12}$-acetate in 10 mg acetonitrile) concentration was prepared in a beaker. The top ~ 90 % portion of this solution was used to assure that no sediment was present. Subsequently, this solution was evaporated over the course of 1 hour in a glass Petri dish. The magnetization sample was obtained by scraping the powder from the dish. Magnetic measurements were acquired in a Quantum Design MPMS-XL SQUID magnetometer.

We note, that the as-produced $Mn_{12}$-acetate displayed magnetic data as seen before by others. Specifically, differences between the zero-field cooled (ZFC) and field cooled (FC) magnetization data display the customary signatures of a blocking temperature at ~ 3K. Furthermore, the ac-susceptibility shows a typical frequency dependence with peaks in the imaginary part at 4.2K (1Hz), 6K (100Hz) and 7.8K (1000Hz). In contrast, novel effects were observed in the magnetization data of the solvent exposed $Mn_{12}$-acetate in both ac-susceptibility, ZFC and FC magnetization measurements. The temperature dependence of the in-phase ($\chi'$) and out-of-phase ($\chi''$) ac-susceptibility for the sample were investigated in an ac field of 3 Oe at different frequencies ranging from 1 Hz to 1000 Hz (Fig. 3). A magnetic transition was observed at ~ 11 K in the $\chi'$ data. As the frequency was increased from 1 Hz to 1000 Hz, an increase in the transition temperature was observed. The inset shows a blow up around the transition temperature. Maxima, at 10.5 K (1 Hz) and 10.8 K (10 Hz), were observed in the $\chi''$ data. Moreover, a frequency dependence of the peaks for 100 Hz and 1000 Hz is also evident at ~ 11.0 K and ~ 11.6 K, respectively, although the data are noisier. A shift of the maxima in the $\chi''$ data from those of crystalline $Mn_{12}$-acetate (e.g. 5.0 K at 10 Hz) [21] is consistent with an increase of the blocking temperature ($T_B$) in our film material.

Fig. 4 shows both ZFC and FC magnetization curves, which are split below $T_B$ ~ 9.6 K in $H_{app}$ = 0.05 T [22]. $T_B$ is shown in the inset of Fig. 4 as a function of applied magnetic fields, $H_{app}$ = 0.001, 0.005, 0.05, 0.3, and 0.5 T, respectively. An increased $T_B$ around 10 K is consistent with AC data. The magnetization of the sample was measured as a function of the applied magnetic field at 1.8 K. A hysteresis loop with a coercive field of ~ 760 Oe was obtained. The low coercive field may be due to a random orientation of the sample.

Considering the fact that the core of $Mn_{12}$-acetate molecules is not affected by acetonitrile, as was shown in a previous report [23], we suggest the following possible reasons for the shifted $T_B$ compared to that of the parent compound: (1) The change in magnetic properties may be the result of structural changes to the $Mn_{12}$-acetate complex, e.g. missing ligands from the $Mn_{12}$-acetate molecules or missing water of crystallization (2) The total spin of the $Mn_{12}$-acetate molecules may have changed in this process. (3) Intermolecular interaction between the $Mn_{12}$-acetate molecules may be enhanced due to missing ligands. At this point we cannot distinguish which of the effects or combination thereof is present in our sample.

## 4. Conclusion

In summary, the deposition of $Mn_{12}$-acetate directly onto a $Si/SiO_2$-substrate using the DAD method yields very smooth, homogeneous thin films. More interestingly, altered magnetic properties were observed from the ac-susceptibility, as well as ZFC and FC magnetization measurements. The blocking temperature of the film material increased to ~ 10 K, indicating that the magnetic anisotropy of the $Mn_{12}$-acetate film may have significantly changed during exposure to acetonitrile. This study is an important step towards the use of molecular magnets in temperature ranges that lend themselves to important applications.

## Acknowledgements

This research was supported by the Texas Advanced Research Program (010366-0038-2001), the Robert A. Welch Foundation(A-1585), the National Science Foundation (DMR-0103455 and NSF-9974899 ), DOE (DOE--DE-FG03-02ER45999), and the Telecommunications and Informatics Task Force (TITF 2001-3) at Texas A&M University.  Use of the TAMU/CIMS Materials Characterization Facility is acknowledged. We thank D. Naugle for helpful discussions

# Figures

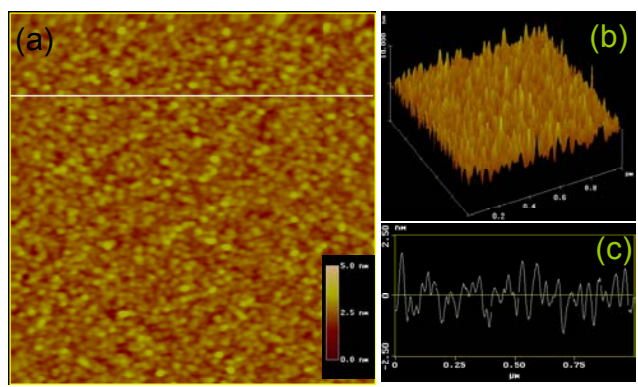

Fig. 1. AFM images of a $Mn_{12}$-acetate thin film formed by the Dip-and-Dry method. (a) Topographical top-view with $1 \times 1~\mu m^2$ scan size. (b) The corresponding 3D image of the same area. (c) Height profile, which is taken at the location indicated by the white line in (a). The height scale bar is shown in the topographical top-view.

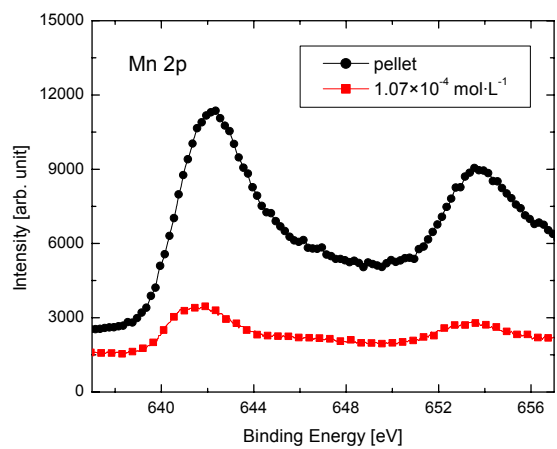

Fig. 2. XPS Spectra of the Mn 2p core level for crystalline $Mn_{12}$-acetate (pellet) and the $Mn_{12}$-acetate thin film.

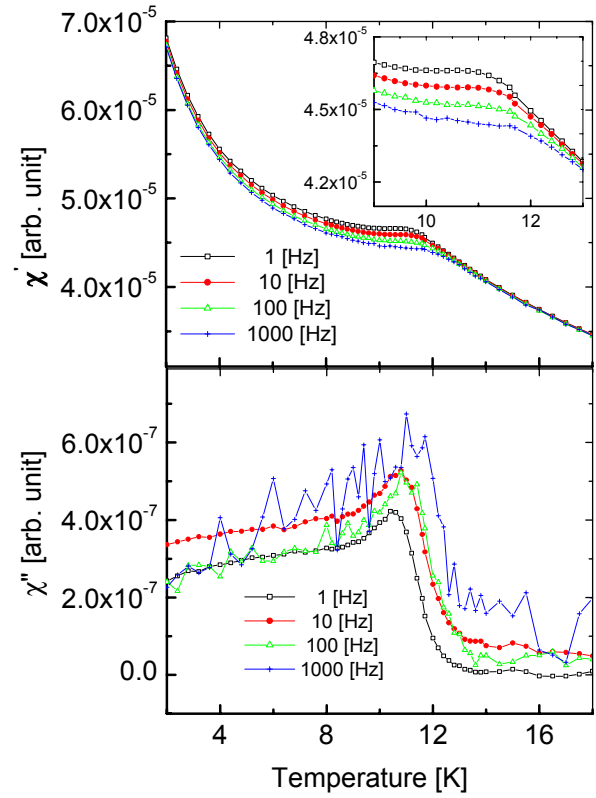

Fig. 3. Temperature dependence of real part ($\chi'$) and imaginary part ($\chi''$) of ac-susceptibility at different frequencies. The inset shows the frequency dependence of the $\chi'$ data around the transition temperature.

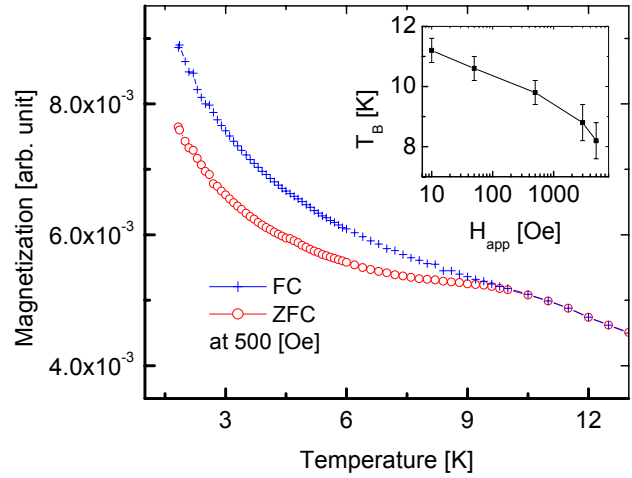

Fig. 4. Zero-field-cooled (ZFC) and field-cooled (FC) magnetization vs temperature curves at $H_{app}$= 500 Oe. The inset shows the blocking temperature ($T_B$) as a function of different applied magnetic fields ($H_{app}$ = 10, 50, 500, 3000, 5000 Oe), respectively. The connected line is a guide for the eye.